\documentclass[aps,physrev,preprint,superscriptaddress,BibTeX,amsmath,amssymb,
]{revtex4-2}

\usepackage{graphicx}
\usepackage{dcolumn}
\usepackage{bm}
\usepackage[pagewise]{lineno}
\usepackage{xcolor}
\begin{document}

\title{\textbf{Dark matter direct search result from InDEx run2 at JUSL} 
}

\author{Susmita Das}
\affiliation{Saha Institute of Nuclear Physics, Kolkata 700064, India}
\affiliation{Homi Bhabha National Institute, Training School Complex, Anushakti Nagar, Mumbai-400094, India}

\author{Mala Das}
\email {mala.das@saha.ac.in}
\affiliation{Saha Institute of Nuclear Physics, Kolkata 700064, India}
\affiliation{Homi Bhabha National Institute, Training School Complex, Anushakti Nagar, Mumbai 400094, India}

\author{Vimal Kumar} 
\affiliation{Saha Institute of Nuclear Physics, Kolkata 700064, India}
\affiliation{Homi Bhabha National Institute, Training School Complex, Anushakti Nagar, Mumbai-400094, India}

\author{Suraj Ali}
\affiliation{Physics Department, Jadavpur University, Kolkata 700032, India }

\author{Nilanjan Biswas}
\affiliation{Saha Institute of Nuclear Physics, Kolkata 700064, India}

\author{Shantonu Sahoo}
\affiliation{Variable Energy Cyclotron Centre, Kolkata 700064, India }

\author{Niraj Chaddha}
\affiliation{Variable Energy Cyclotron Centre, Kolkata 700064, India }


\begin{abstract}
The Indian Dark matter search Experiment (InDEx) has been initiated at Jaduguda Underground Science Laboratory (JUSL) to explore the low mass region of dark matter. The detectors used by InDEx are superheated droplet detectors with active liquid C\(_2\)H\(_2\)F\(_4\). The run1 of InDEx was with 2.47 kg-days of exposure at a threshold of 5.87 keV. In the present work, the run2 of InDEx, the detectors were set at 1.95 keV thresholds with an active liquid mass of 70.4 g. For a runtime of 102.48 days, the experimental results set constraint on spin-independent cross section at \((1.55+(0.62-0.32)_{stat}+(+0.03-0.02)_{sys}) \times 10^{-40}\) cm\(^2\) (90\% C.L.) and spin-dependent limit on F at \((7.97+(+3.44-1.78)_{stat}+(+0.15-0.18)_{sys}) \times 10^{-38}\) cm\(^2\) (90\% C.L.) for 20.4 GeV/c\(^2\) and 21.0 GeV/c\(^2\) weakly interacting massive particle (WIMP) mass respectively. There is a shift of the most sensitive WIMP mass towards the lower region and an improvement of the sensitivity limit over the InDEx run1. 
\end{abstract}

\maketitle


\section{\label{sec:level1}Introduction}

The existence of the massive, nonluminous, nonrelativistic and nonbaryonic dark matter (DM) as a major constituent of the Universe has been confirmed by several astrophysical and cosmological observations. Observational evidences like the rotation curve of the Galaxy, gravitational lensing, and anisotropy in the cosmic microwave background of the Universe have verified the existence of dark matter \cite{damico2009darkmatterastrophysics,Roszkowski_2018}. Among the predicted DM candidates, e.g., neutron stars, black holes, massive astrophysical compact halo object, neutrino, weakly interacting massive particles (WIMPs), axions, etc., the WIMPs are one of the most favored candidates. Studies are being conducted worldwide to find dark matter using direct detection, indirect detection, and collider searches; however, this article focuses on the direct detection experiment. Current direct detection experiments on WIMP searches are predominantly focused on three types of WIMP interaction: DM-nucleus scattering, DM-nucleus scattering considering the Migdal effect, and DM-electron scattering \cite{PhysRevLett.129.221301}. In DM-nucleus interaction, several experiments have been conducted (and are ongoing) with different types of detectors such as noble gas (XENON, LUX-ZEPLIN, PandaX, DarkSide-50), superheated liquid (PICASSO [C\(_4\)F\(_{10}\)], PICO [C\(_3\)F\(_8\)], SIMPLE [C\(_2\)ClF\(_5\)], MOSCAB [C\(_3\)F\(_8\)], COUPP [CF\(_3\)I]) and crystal (SuperCDMS, CDEX) to probe the WIMPs in GeV to TeV mass range \cite{PhysRevLett.131.041003,PhysRevLett.131.041002, PhysRevLett.130.261001,ARNAUD201854,AGNES2015456,BEHNKE201785,PhysRevD.100.022001, PhysRevLett.108.201302,Antonicci2017,fustin2024darkmatterlimitscoupp,PhysRevD.107.122003,PhysRevD.99.062001,PhysRevLett.120.241301}. To explore the lower mass region under the DM-nucleus interaction paradigm, the analysis with the Migdal effect has been done by a few experiments such as CDEX, LUX, and CRESST \cite{PhysRevLett.129.221301,PhysRevLett.123.161301,PhysRevLett.122.131301,Angloher2017}. The experiments such as CDEX, SENSEI, DAMIC, EDELWEISS, SuperCDMS, XENON, PandaX and DarkSide \cite{PhysRevLett.129.221301,PhysRevLett.134.011804,PhysRevLett.123.181802,PhysRevLett.125.141301,PhysRevD.102.091101,PhysRevD.96.043017,PhysRevLett.123.251801,PhysRevLett.126.211803,PhysRevLett.121.111303} have adopted DM-electron scattering which has contributed to lowering the WIMPs detection mass region to the MeV range. Under DM-nucleus scattering consideration, the lower mass (MeV) region of WIMPs has been explored by NEWS-G (noble gas) and CRESST-III (Si crystal) experiments \cite{ARNAUD201854,PhysRevD.107.122003}. The MeV mass region of WIMPs can also be probed by superheated liquid detectors containing light target elements \cite{PhysRevD.101.103005}. The InDEx (Indian Dark matter search Experiment) detectors are based on the superheated liquid, C\(_2\)H\(_2\)F\(_4\) and focused on DM-nucleus elastic scattering \cite{PhysRevD.101.103005}. To understand the coupling between WIMPs and standard model particles in terms of spin-independent (SI) and spin-dependent (SD) cross sections, both spin zero (C) and nonzero (F, H) target elements are chosen. The presence of H as a target element makes the detector suitable to explore the low mass (up to MeV) region of WIMPs at a very low detector threshold  \cite{PhysRevD.101.103005}. The possibility of background rejection depending upon the threshold energy of the detector acts as a major advantage for this kind of detector using superheated liquid. The physics runs of InDEx are commissioned at Jaduguda Underground Science Laboratory (JUSL), Jharkhand at a depth of 555 m rock overburden or 1604 m water equivalent \cite{GHOSH2022102700}. In run1, an exposure of 2.47 kg-days was explored at the threshold of 5.87 keV by operating the detector at the ambient temperature \cite{kumar2025resulttetrafluoroethanec2h2f4superheated}. In the present work, run2, two superheated droplet detectors of C\(_2\)H\(_2\)F\(_4\), having in total 70.4 g of active liquid, were set at a threshold of 1.95 keV within a temperature-controlled stainless steel (SS) container. The experiment ran for 102.48 days resulting in an exposure of 7.2 kg-days. In the following sections, the experimental methods, estimation of the count rate for the background and WIMPs followed by the exclusion plots are discussed. Finally, the article ends with a discussion of the experimental observations.

\section{\label{sec:level2}Experimental Methods}

\begin{figure}
\includegraphics[width=8.5cm]{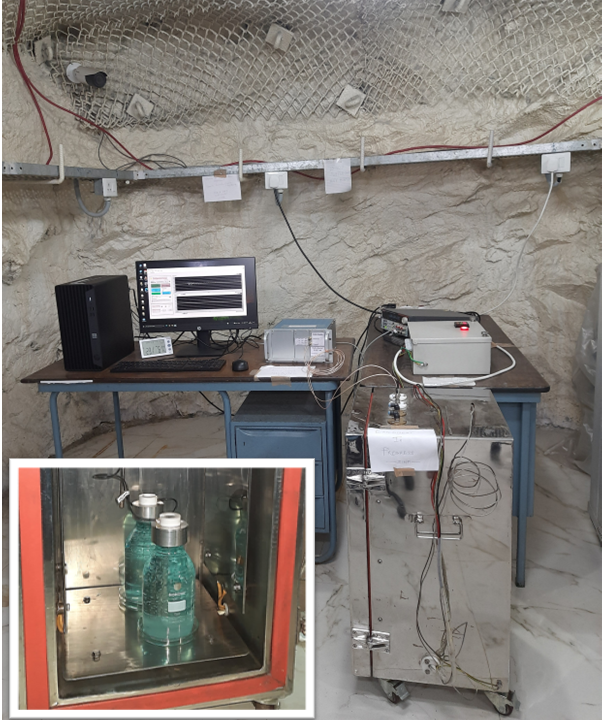}%
\caption{\label{fig:1}Experimental setup of InDEx run2 at JUSL and the detectors (inset).}
\end{figure}

\begin{figure}
\includegraphics[width=8.5cm]{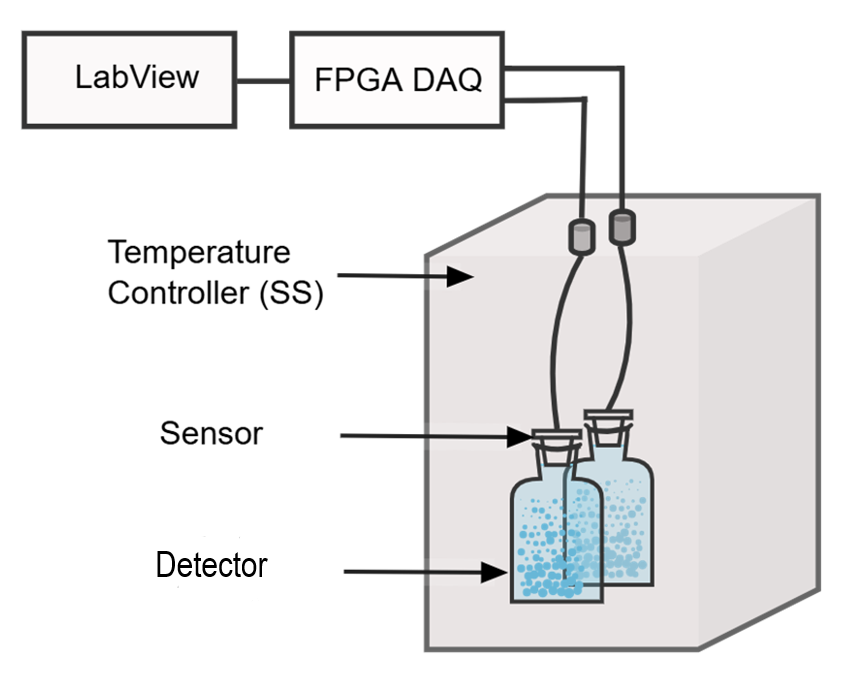}%
\caption{\label{fig:2}Block diagram of the experimental setup.}
\end{figure}

The superheated droplet detectors of C\(_2\)H\(_2\)F\(_4\) were fabricated in a pressure reactor (make: Amar Equipment’s Pvt. Ltd.) at the laboratory of Saha Institute of Nuclear Physics (SINP). The fabricated emulsion was transferred in the borosilicate glass containers. The loading factor of active liquid C\(_2\)H\(_2\)F\(_4\) in viscoelastic gel matrix was 7.6\%. The active liquid mass in one of the detectors (detector1) was 26.4 g in 300 ml emulsion and the other (detector2) was 44.0 g in 500 ml emulsion. To achieve the required thresholds of 1.95 keV in run2, two detectors were installed within a temperature-controlled SS system (make: Bhargab Engineering Pvt. Ltd.), and the operating temperature was set to \((35\pm0.1)\)\(^{\circ}\)C. Each detector was connected with a sensor (make: Physical Acoustic India Pvt. Ltd) on top of it. The sensors were piezoelectric acoustic sensors having a case material made of stainless steel and a face material made of ceramic with frequency in the ultrasound range. The ceramic face was in touch with the aquasonic gel in the detector. The output from the sensors was connected to the FPGA-based data acquisition system and LabView software \cite{SAHOO2021165457}. In LabView, for the data acquisition, the amplitude threshold was set at 150 mV above the noise level. The experimental setup of InDEx run2 at JUSL is shown in Fig. \ref{fig:1}. and the block diagram is shown in Fig. \ref{fig:2}. The calibration run was done with a \(^{241}\)AmBe (10 mCi) neutron source at the same temperature (35\(^{\circ}\)C) as that of the WIMP run at JUSL.

\section{\label{sec:level3}Data selection}

\begin{figure}
\includegraphics[width=8.5cm]{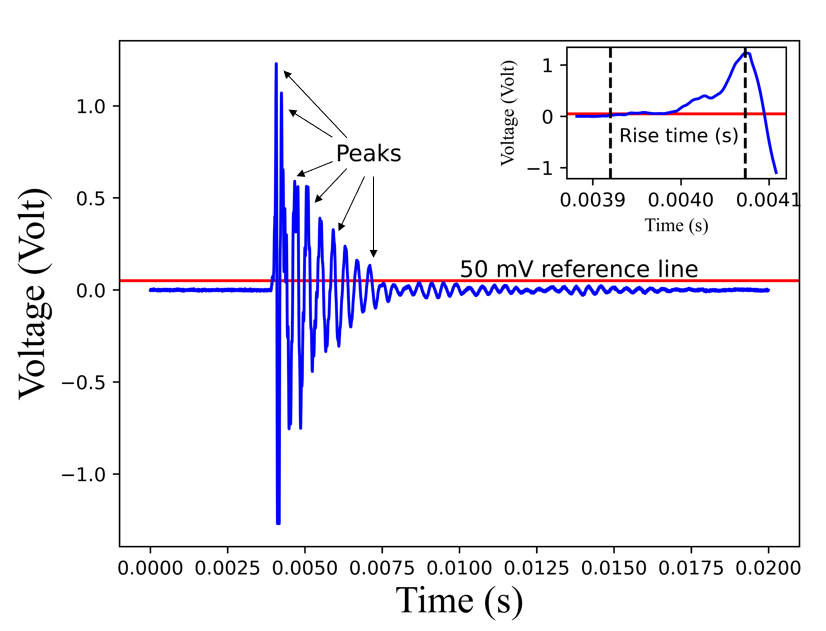}%
\caption{\label{fig:3}Typical signal from the neutron calibration run. Inset: figure rise time.}
\end{figure}

In the experimental setup for run2, two detectors were independently connected with two sensors and the DAQ system. The recorded data were stored separately in two channels for the two detectors. Events from the two detectors were added to get the total event. In the selection of run2 data, the bubble nucleation events were separated from the noise events by applying several selection cuts on the parameters as constructed for each of the events. To eliminate the external noise events, the parameters such as signal rise time and number of peaks (N\(_{peak}\)) from the calibration data have been constructed. Signal rise time is the time required to reach the peak of the acoustic signal above the 50 mV reference line. The N\(_{peak}\) is the number of peaks present in a signal above the 50 mV reference line. The choice of 50 mV is mainly due to the baseline fluctuation (30-40 mV) of a signal during the calibration experiment with \(^{241}\)AmBe for the FPGA-based DAQ. A typical calibration signal with rise time and peaks is shown in Fig. \ref{fig:3}. The chosen cut values and their effectiveness are presented in Table \ref{tab:table 1}.

\begin{table}
\caption{\label{tab:table 1} Different cuts applied on parameters for signal and noise separation.}
\begin{ruledtabular}
\begin{tabular}{lcdr}

\textrm{Parameters}&
\textrm{Calibration experiment}&
\textrm{InDEx run2}\\
\colrule
Triggered events & 220 & 108 \\
Rise time $>$ 10 \(\mu\)s. & 220 & 95 \\
N\(_{peak}\) $>$ 20 & 220 & 92 \\
S-checking & 220 & 85\\
\end{tabular}
\end{ruledtabular}
\end{table}

\begin{figure}
\includegraphics[width=8.5cm]{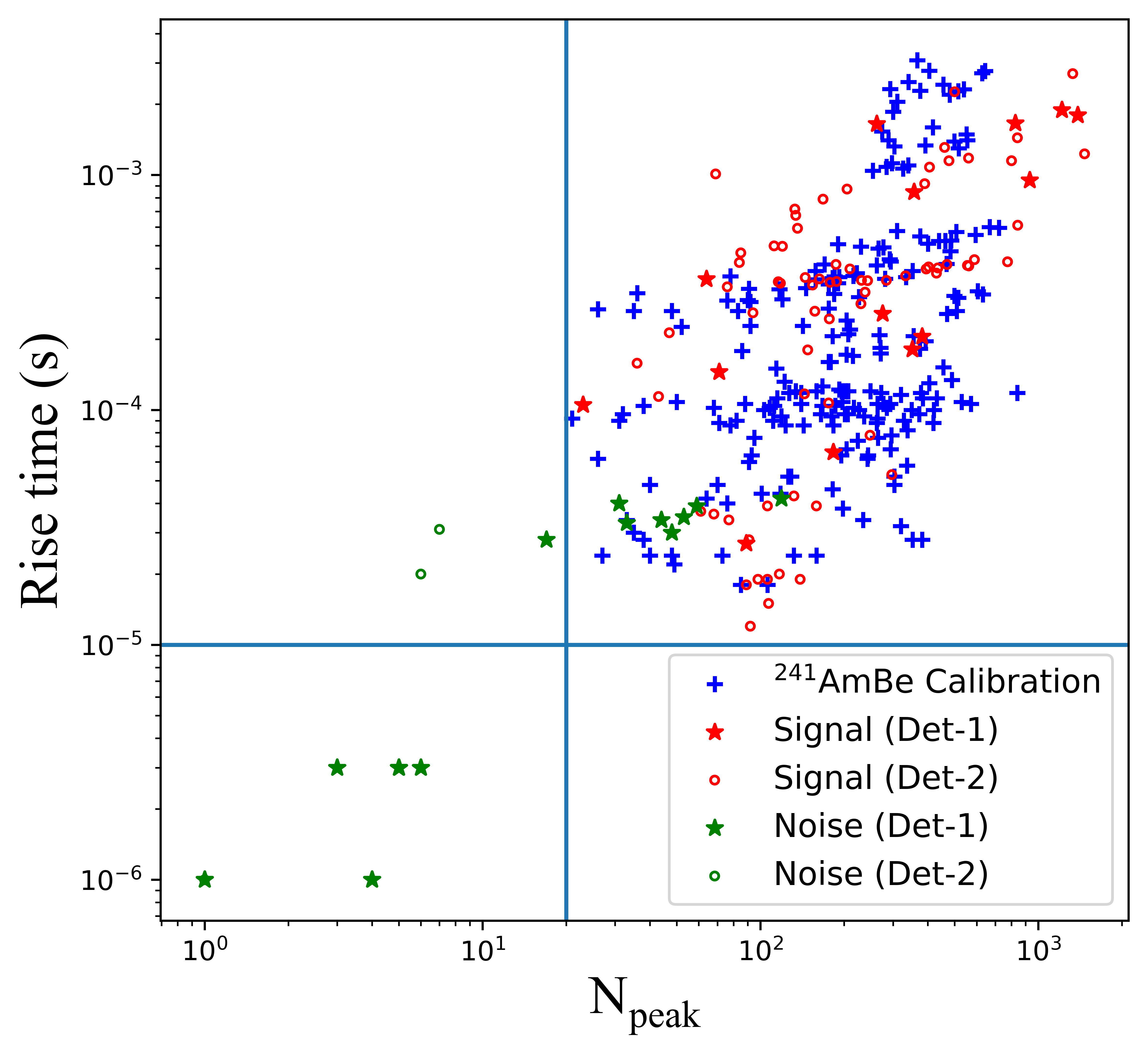}
\caption{\label{fig:4}Rise time vs N\(_{peak}\) distribution for the triggered events of InDEx run2 and neutron calibration. The vertical and horizontal lines correspond to the cut values.}
\end{figure}

Applying these cuts on the run2 dataset, 92 events were collected as acoustic signals from nucleation events. Further, by applying the symmetry (S) cut, 85 acoustic signal events have been selected by eliminating all noises. The S parameter passes the signals having both positive and negative amplitudes which is the case of the actual bubble nucleated events recorded by the acoustic sensor and rejects those having only one-sided amplitudes. The distribution of rise time vs N\(_{peak}\) for the calibration and run2 signals for individual detectors is shown in Fig. \ref{fig:4}.

\section{\label{sec:level4}Response to the backgrounds}

The major backgrounds present at JUSL are neutrons and \(\gamma\)-rays. At the operating temperature of 35 \(^{\circ}\)C, the InDEx detectors are insensitive to \(\gamma\)-rays \cite{SAHOO2021165450}, hence neutrons are the dominant background for run2. Earlier publication of our group shows that C\(_2\)H\(_2\)F\(_4\) starts to  respond to \(\gamma\)-rays above \((38.5\pm1.4)\) \(^{\circ}\)C \cite{SAHOO2021165450, SAHOO201944}. To verify that additionally, in the present work, the detector was irradiated with \(^{137}\)Cs (5 mCi) at 35 \(^{\circ}\)C for about 1 h and no excess of count over the background was observed for this measurement.

The sensitivity of C\(_2\)H\(_2\)F\(_4\) to neutrons at 1.95 keV thresholds has been calculated by considering the distances traveled by C, F, and H nuclei within C\(_2\)H\(_2\)F\(_4\) by satisfying the bubble nucleation condition \cite{PhysRevD.101.103005}. The ranges have been calculated by the “Stopping and Range of Ions in Matter” (SRIM) software package \cite{Zeigler}. The calculation leads to 7 keV and 11 keV of incident neutron energy that can trigger bubble nucleation at 35 \(^{\circ}\)C for C and F respectively. The ranges of C and F for 7 keV and 11 keV neutron energies are 14.12 nm and 11.5 nm respectively. In the case of H, it is insensitive to neutrons at a detector threshold energy of 1.95 keV (35 \(^{\circ}\)C) \cite{PhysRevD.101.103005}. 

As stated earlier at 1.95 keV (35 \(^{\circ}\)C), the main background is neutrons, and therefore an estimation of the detector count rate from the background neutrons is needed. The measurement of neutron flux, the simulation of the spectrum and the flux for both the radiogenic and cosmogenic neutrons inside JUSL have already been done earlier \cite{GHOSH2022102700}. In the setup of InDEx run2, two detectors were placed inside a temperature-controlled system made of 5-cm-thick SS of dimension 69 cm x 40.5 cm x 5 cm. To estimate the neutron flux (\(\phi(E_n)\)) in the present setup, the FLUKA \cite{fluka} and flair \cite{flair} simulation toolkits with their source routine were used. The neutrons having an energy spectrum as mentioned in the simulation result of Ref. \cite{GHOSH2022102700} are allowed to pass through a 5-cm-thick SS slab before to incident on the detector. During the flux calculation, the measured value of radiogenic neutron flux with neutron energy \(\geq\) 100 keV and the simulated value of cosmogenic neutron flux of neutron energy \(\geq\) 1 MeV is considered as available in the literature \cite{GHOSH2022102700}. The event rate in the detector for this setup has been estimated using Eq. (\ref{eq:one}) \cite{SAHOO2021165450}, as follows:

\begin{eqnarray}
R(E_n,T) = {\phi(E_n)V_l{\sum^{N}_{i}}\epsilon^i(E_n,T)N^i\sigma_n^i(E_n) }
\label{eq:one}
\end{eqnarray}
where,\\
\(\phi(E_n)\) = neutron flux at an energy \(E_n\) incident on the detector active volume,\\
\(V_l\) = detector active liquid volume, \\
\(N^i\) = atomic number density of \(i^{th}\) element of active liquid,\\
\(\sigma_n^i(E_n )\) = cross section of \(i^{th}\) element of active liquid at energy \(E_n\),\\
\(\epsilon^i\) = detector efficiency of \(i^{th}\) element of active liquid. 

The cross sections \(\sigma_n (E_n )\) are calculated as the average cross section from the NNDC \cite{NNDC} dataset for elastic scattering and the detector efficiency (\(\epsilon\)) is considered from Ref. \cite{PhysRevD.101.103005}. For the active liquid C\(_2\)H\(_2\)F\(_4\), the event rate has been estimated for radiogenic neutrons as \((1.91\pm0.10_{stat}\pm 0.03_{sys})\times 10^{-7}\) s\(^{-1}\) g\(^{-1}\) and for cosmogenic neutrons as \((2.38\pm0.09_{stat}\pm 0.04_{sys})\times10^{-11}\) s\(^{-1}\) g\(^{-1}\). Here the statistical errors are due to the statistical uncertainties present in the simulated fluxes, \(\phi(E_n)\), the systematic errors are due to the error in volume measurement, \(V_l\), and the error in efficiency, \(\epsilon\). The error in efficiency is due to temperature fluctuation during the measurement. Both the statistical and systematic errors in neutron event rate due to radiogenic (5.2\% and 1.6\%) and cosmogenic (3.8\% and 1.7\%) cases are well below 10\%. Since the experimental count rate is within the expected count rate from the background neutrons, the main contributing background is from neutrons and hence the contribution from any other background (e.g. alpha) is not considered here. In order to verify the simulation, the expected count rate for the calibration experiment has been calculated for \(^{241}\)AmBe spectrum using FLUKA. The expected rate for the calibration run is found to be \((2.66\pm0.01_{stat}\pm 0.36_{sys})\times10^{-3}\) s\(^{-1}\) g\(^{-1}\) which is close to the experimental event rate \((2.03\pm0.14_{stat}\pm 0.28_{sys})\times10^{-3}\) s\(^{-1}\) g\(^{-1}\) for calibration. The higher systematic error in the calibration run is due to the smaller active liquid volume of the detector and a temperature precision of \(\pm1\) \(^{\circ}\)C. The background measurement at JUSL shows the presence of both neutrons and \(\gamma\)-rays. But at 1.95 keV thresholds, C\(_2\)H\(_2\)F\(_4\) SLD is insensitive to \(\gamma\)-rays. Therefore the only possible background at 1.95 keV which mainly comes from the neutrons is known with negligible uncertainty from all sources.

\section{\label{sec:level5}WIMP search result}

The basic purpose of WIMP search experiments is to search for the excess over the estimated background events. In that case, statistical inferences based on observed data are important. At present for run2 of InDEx, the profile likelihood ratio method has been considered for statistical interpretation of the experimental result \cite{article,Baxter2021}. The likelihood function \textit{L} for counting experiments having a parameter of interest (\(\mu\)) with known background can be defined by Eq. (\ref{eq:two}) \cite{article}, where \textit{n}, \textit{s} and \textit{b} represent the numbers of observed events, expected signal events and the expected background events respectively. In the present case, the \(\mu\) has been considered as the WIMP-nucleon SI cross section and \textit{s} is the number of WIMP events per unit WIMP-nucleon SI cross section \cite{PhysRevD.101.103005}. The \textit{b} is taken as the mean value of expected events from the known neutron background. For a positive parameter of interest (as the cross section here), the test statistic \(\tilde{t_\mu}\) \cite{article,Baxter2021} takes the form of Eq. (\ref{eq:three}) with \(\hat\mu\) that maximizes L, as follows:

\begin{eqnarray}
L(\mu) = \frac{(\mu s+b)^n}{n!} e^{-(\mu s+b)}
\label{eq:two}
\end{eqnarray}

\begin{eqnarray}
\tilde{t_\mu} =\begin{cases}
      -2log\frac{L(\mu)}{L(\hat\mu)} & \hspace{0.5cm}  \hat\mu\geq0, \\
      -2log\frac{L(\mu)}{L(0)} & \hspace{0.5cm} \hat\mu < 0. \\
      
    \end{cases} 
\label{eq:three}
\end{eqnarray}

The \textit{s} has been calculated by normalizing the expected event rate, \(R_{exp}\) with the experimental exposure and WIMP-nucleon SI cross section. The \(R_{exp}\) due to WIMP can be estimated by Eq. (\ref{eq:four}) \cite{PhysRevD.101.103005}, as follows:

\begin{eqnarray}
R_{exp} = \sum_{i}\xi_i\int_{E_{R,th}^{(i)}}^{E_{R,max}^{(i)}} dE_R \epsilon _i{(E_R)}\left(\frac{dR}{dE_R}\right)_i
\label{eq:four}
\end{eqnarray}

where, \(\xi_i\) is the mass fraction of target element \textit{i}, \(\frac{dR}{dE_R}\)  is the differential recoil rate for a detector, \(\epsilon_i{(E_R)}\) is the bubble nucleation efficiency, \(E_{R,th}^{(i)}\) is the recoil energy threshold for bubble nucleation by nuclei of an element \textit{i} and \(E_{R, max}^{(i)}\) is the maximum recoil energy a nucleus of an element \textit{i} can receive due to scattering with a WIMP.
The standard halo parameters, \(\rho_D\)  = 0.3 GeV c\(^{-2}\) cm\(^{-3}\), \(v_{esc}\) = 540 km s\(^{-1}\), \(v_E\) = 232 km s\(^{-1}\), and \(v_0\) = 220 km s\(^{-1}\) have been used for this calculation with the WIMP-nucleon SI cross section taken as 1 pb \cite{PhysRevD.101.103005}.
To verify the presence of a new signal, the background-only null hypothesis has been tested with \(\mu\) = 0, and the \textit{p}-value (\(p_0\)) has been calculated. The \textit{p}-value (\(p_0\)) is the probability of observing the test statistic \(\tilde{t_\mu}\) under the null hypothesis ( \(\mu\) = 0) which can then be inferred as a level of disagreement between the experimental data and background hypothesis. Under \(\mu\) = 0 condition, \(\tilde{t_\mu}\) becomes to \(\tilde{t_0}\), as defined in Eq. (\ref{eq:five}) and \(p_0\) can be calculated by Eq. (\ref{eq:six}) \cite{article}, as follows:

\begin{eqnarray}
\tilde{t_0} =\begin{cases}
      -2log\frac{L(0)}{L(\hat\mu)} & \hspace{0.5cm}  \hat\mu\geq0, \\
      0 & \hspace{0.5cm}  \hat\mu < 0, \\
      \end{cases} 
\label{eq:five}
\end{eqnarray}

\begin{eqnarray}
p_0 = 1-\phi\left(\sqrt{\tilde{t_0}} \right),   
\label{eq:six}
\end{eqnarray}

where \(\phi\)  is the cumulative distribution of a standard Gaussian. The \textit{p}-value below  \(p_{3\sigma}\)=\(1.4 \times 10^{-3}\) can be claimed as "evidence" and the calculated \textit{p}-value for InDEx is 0.50. Hence an exclusion limit can be set on the parameter space of WIMP-nucleon SI cross section \(\sigma_{\chi n}^{SI}\) vs. WIMP mass. Here the test statistic \(\tilde{t_\mu}\), as recommended in Ref. \cite{Baxter2021} is used to calculate the upper limit on \(\mu\) at 90\% confidence levels (C.L.). For a specific WIMP mass, the \textit{p}-value has been calculated first by considering a hypothesized value of \(\mu\) using Eqs. (\ref{eq:seven}) and (\ref{eq:eight}) where the value of \(\sigma\) has been calculated using Eq. (\ref{eq:nine}) assuming the Wald approximation for \(\tilde{t_\mu}\) \cite{article}. The same procedure has been repeated over a large set of \(\mu\) values to find the specific \(\mu\) for which the \(p_\mu\) satisfies the 90\% C.L., as follows:

\begin{eqnarray}
p_\mu = 1-F\left(\tilde{t_\mu}|\mu \right),   
\label{eq:seven}
\end{eqnarray}
where
\begin{eqnarray}
F\left(\tilde{t_\mu}|\mu \right) =\begin{cases}
      2\phi \left(\sqrt{\tilde{t_\mu} }\right)-1   & \hspace{0.5cm}  \tilde{t_\mu}\leq\mu^2/\sigma^2, 
      \\
      \phi \left(\sqrt{\tilde{t_\mu} }\right)+\phi \left(\frac{\tilde{t_\mu}+\mu^2/\sigma^2}{2\mu/\sigma} \right)-1 & \hspace{0.5cm} \tilde{t_\mu}>\mu^2/\sigma^2, \\
      \end{cases}  
\label{eq:eight}
\end{eqnarray}
and
\begin{eqnarray}
\tilde{t_\mu} =\begin{cases}
      \frac{\mu^2-2\mu\hat\mu}{\sigma^2}   &\hspace{0.5cm}  \hat\mu<0, 
      \\
      \frac{(\mu-\hat\mu)^2}{\sigma^2}
       & \hspace{0.5cm}  \hat\mu\geq0. \\
      \end{cases}  
\label{eq:nine}
\end{eqnarray}

To calculate the WIMP-proton SD cross section \(\sigma_p^{SD}\) for F, the expected events from backgrounds and WIMPs are normalized with the active mass of F. The 90\% C.L. upper limit on the WIMP-nucleon SI cross section for F, \(\sigma_{\chi n}^{SI(F)}\) has been calculated using the method as described earlier. It is then converted to \(\sigma_p^{SD}\) using the Eqs. (\ref{eq:ten}) and (\ref{eq:eleven}) as stated below \cite{BEHNKE201785,PhysRevD.101.103005}: 

\begin{eqnarray}
\sigma_{\chi n}^{SI(F)}(0) = \sigma_{\chi n}^{SI(F)} \frac{(1+\frac{m_\chi}{m_n})^2}{(1+\frac{m_\chi}{m_A})^2} A^2
\label{eq:ten}
\end{eqnarray}

\begin{eqnarray}
\sigma_p^{SD} = \sigma_{\chi n}^{SI(F)}(0) \left(\frac{\mu_p}{\mu_F}\right)^2
\frac{C_p^{SD}}{C_{p(F)}^{SD}}
\label{eq:eleven}
\end{eqnarray}

where, \(\sigma_{\chi n}^{SI(F)} (0)\) = WIMP-nucleus SI cross section (zero momentum) for F, \(\sigma_{\chi n}^{SI(F)}\) = WIMP-nucleon SI cross section for F, \(m_\chi\) = WIMP mass, \(m_n\) = nucleon mass (proton or neutron), \(m_A\) = nuclear mass, A = mass numbers of F, \(\sigma_p^{SD}\) = spin dependent cross section for free protons of F, \(\mu_{p(F)}\) = WIMP-proton (fluorine) reduced mass, \(\frac{C_p^{SD}}{C_{p(F)}^{SD}}\) = 1.285 \cite{BEHNKE201785}.

\begin{figure}
\includegraphics[width=8.5cm]{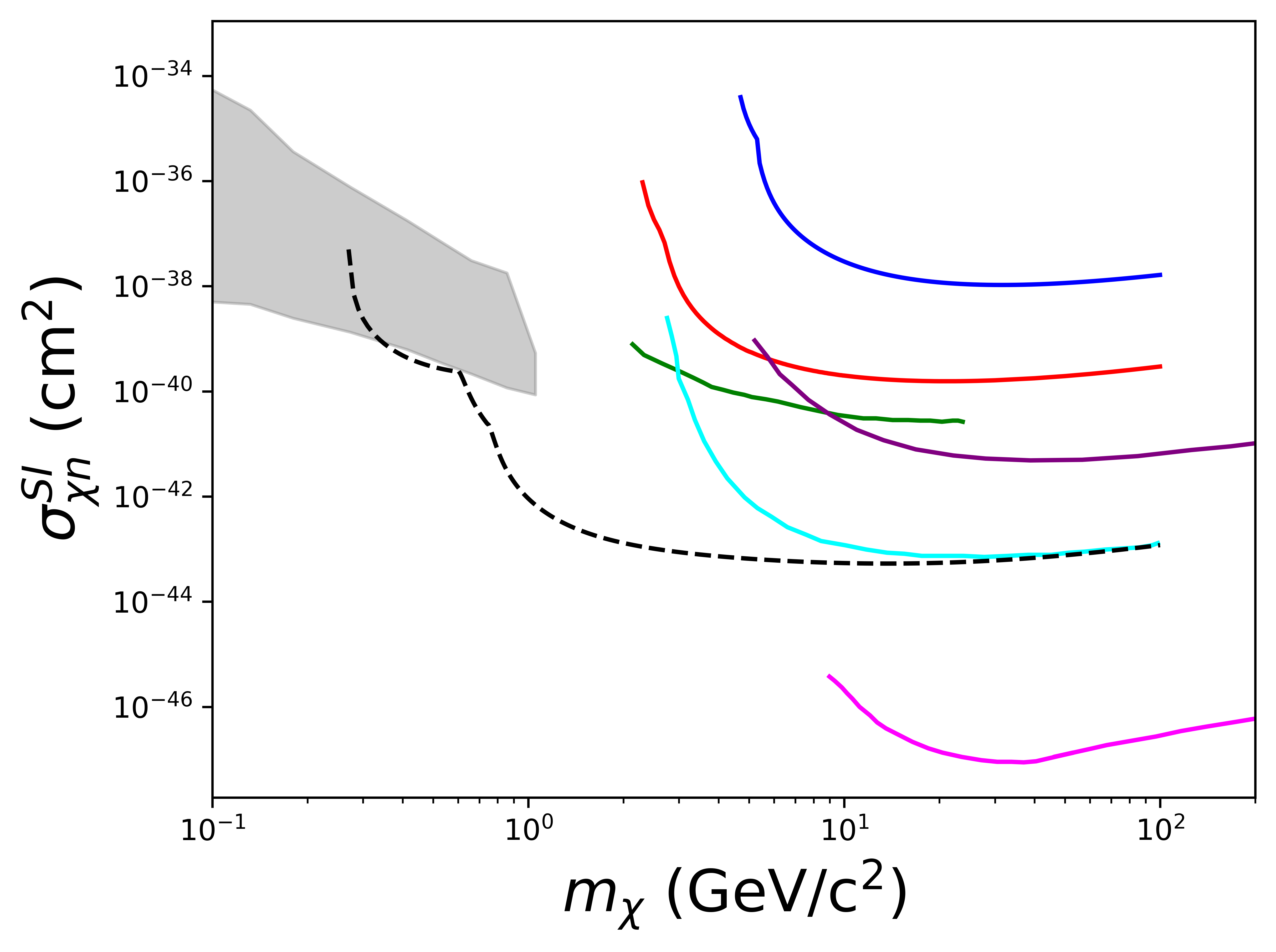}
\caption{\label{fig:5}Present (red solid line) and projected sensitivity (black dashed line) of InDEx along with the results from SIMPLE \cite{PhysRevLett.108.201302} (purple solid line), PICASSO \cite{BEHNKE201785} (green solid line), PICO \cite{PhysRevD.100.022001} (aqua solid line), LUX-ZEPLIN \cite{PhysRevLett.131.041002} (magenta solid line), InDEx run1 \cite{kumar2025resulttetrafluoroethanec2h2f4superheated} (blue solid line) and the theoretically predicted region \cite{essig2023snowmass2021cosmicfrontierlandscape,battaglieri2017cosmicvisionsnewideas} (gray contour region) on WIMP-nucleon SI cross section with 90\% C.L.}
\end{figure}

\begin{figure}
\includegraphics[width=8.5cm]{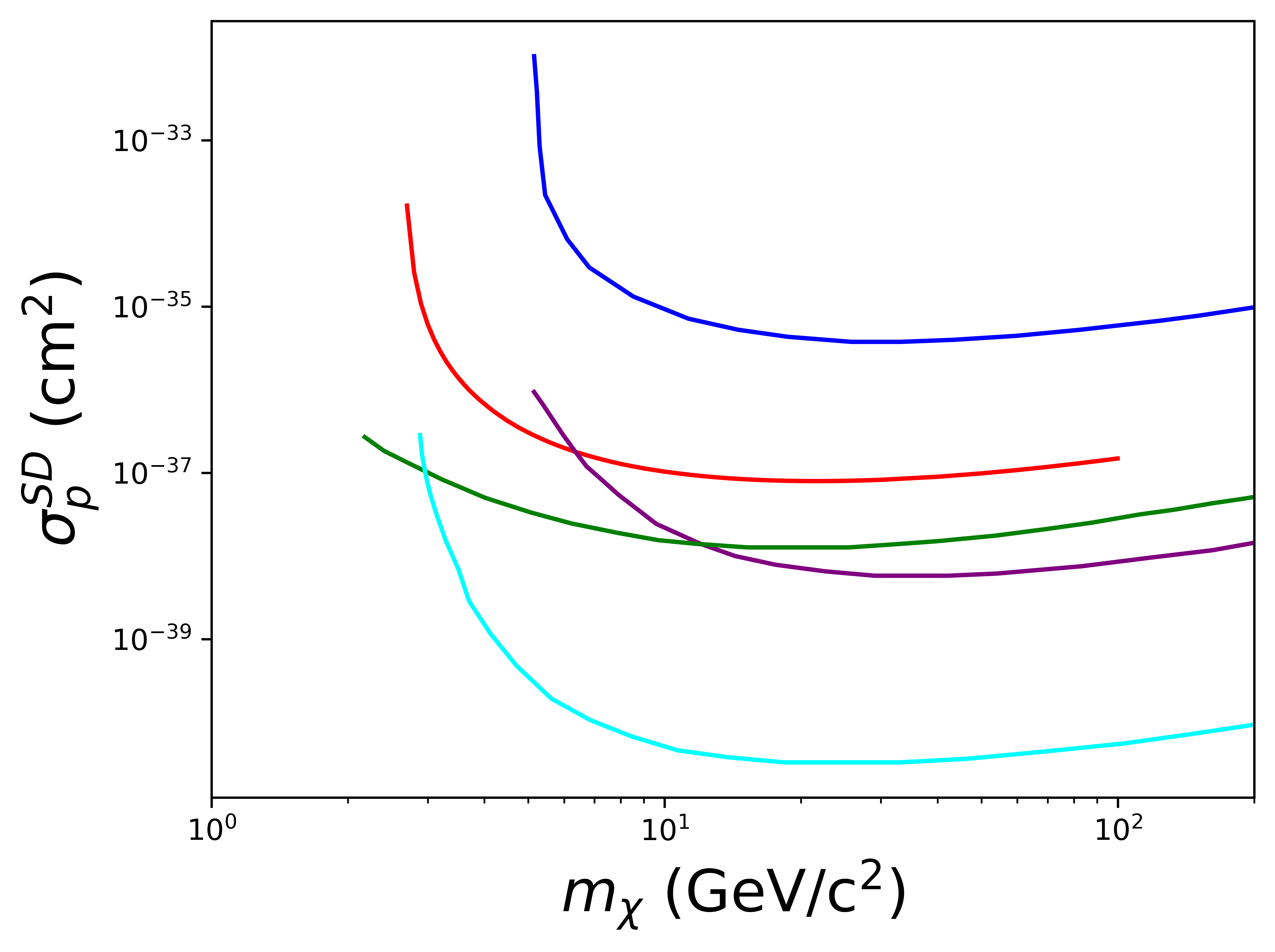}
\caption{\label{fig:6}Present (red solid line) sensitivity of InDEx along with the results from SIMPLE \cite{PhysRevLett.108.201302} (purple solid line), PICASSO \cite{BEHNKE201785} (green solid line), PICO \cite{PhysRevD.100.022001} (aqua solid line) and InDEx run1 \cite{kumar2025resulttetrafluoroethanec2h2f4superheated} (blue solid line) on WIMP-proton SD cross section with 90\% C.L.}
\end{figure}

To get the exclusion plot on parameter space, the whole process has been followed for each mass of WIMPs. The best limit on spin-independent cross section comes for \(\sigma_{\chi n}^{SI}\)= \((1.55+(0.62-0.32)_{stat}+(+0.03-0.02)_{sys}) \times 10^{-40}\) cm\(^2\) (90\% C.L.) at 20.4 GeV/c\(^2\) WIMP mass. Similarly, the spin-dependent upper limit on F shows \(\sigma_p^{SD}\) = \((7.97+(+3.44-1.78)_{stat}+(+0.15-0.18)_{sys}) \times 10^{-38}\) cm\(^2\) (90\% C.L.) at 21.0 GeV/c\(^2\). The errors in the cross section are due to the statistical and systematic uncertainty present in the experimental results. The sensitivity plots for spin-independent and spin-dependent cross sections of WIMP interaction at 90\% C.L. with WIMP mass are shown in Figs. \ref{fig:5} and \ref{fig:6} respectively. The future projected sensitivity of InDEx at 0.19 keV (55\(^{\circ}\)C operating temperature) threshold for 1000 kg-days of exposure is shown in Fig. \ref{fig:5} for the zero background consideration. The limits from other experiments e.g. PICASSO \cite{BEHNKE201785}, SIMPLE \cite{PhysRevLett.108.201302}, and PICO \cite{PhysRevD.100.022001} that use superheated liquids and the leading experiments in SD, SI e.g. PICO \cite{PhysRevD.100.022001} and LUX-ZEPLIN \cite{PhysRevLett.131.041002} along with the theoretically predicted region \cite{essig2023snowmass2021cosmicfrontierlandscape,battaglieri2017cosmicvisionsnewideas} are shown in Figs. \ref{fig:5} and \ref{fig:6}.

\section{\label{sec:level6}Discussion}
The InDEx run2 was carried out at JUSL with two superheated droplet detectors at 1.95 keV thresholds for an exposure of 7.2 kg-days. In this run no WIMP signal was observed above the background and an SD limit on F of \((7.97+(+3.44-1.78)_{stat}+(+0.15-0.18)_{sys}) \times 10^{-38}\) cm\(^2\) and an SI limit of \((1.55+(0.62-0.32)_{stat}+(+0.03-0.02)_{sys}) \times 10^{-40}\) cm\(^2\) for C\(_2\)H\(_2\)F\(_4\) was set at 21.0 GeV/c\(^2\) and 20.4 GeV/c\(^2\) respectively at 90\% C.L. In DM-nucleus elastic scattering, the most stringent limit until now has been set by the LUX-ZEPLIN experiment with liquid xenon dual-phase TPC.  With 0.9 ton-yr of exposure and WIMP sensitivity above 9 GeV/c\(^2\), LUX-ZEPLIN has achieved spin-independent cross section limit at 9.2 x 10\(^{-48}\) cm\(^2\) for a WIMP mass 36 GeV/c\(^2\) \cite{PhysRevLett.131.041002}. In the spin-dependent sector, PICO (bubble chamber of C\(_3\)F\(_8\)) has the most sensitive limit of 3.2 x 10\(^{-41}\) cm\(^2\) at WIMP mass 25 GeV for an exposure of 1404 kg-days \cite{PhysRevD.100.022001}. The main advantage of InDEx is the presence of H in active liquid C\(_2\)H\(_2\)F\(_4\) which makes it suitable to explore the lower-mass region of WIMPs. At present threshold of InDEx run2, H being insensitive to WIMP interaction, the lowest mass that can be explored is for C from a WIMP mass of 2.3 GeV/c\(^2\). Experiments such as PICO (bubble chamber of C\(_3\)F\(_8\)) and CDEX-10 (Ge crystal) lie in the mass region of above 2 GeV/c\(^2\) of WIMPs. The projected sensitivity of InDEx run2 in Fig. \ref{fig:5} indicates that at the threshold (e.g. 0.19 keV) when H becomes sensitive, InDEx has the potential to explore the MeV mass range of WIMPs. The background of the detector can be reduced by using passive shielding for neutrons, purifying the detector material, controlling the radon in the environment etc. All these processes are in progress and in the near future InDEx would be able to run with the larger detector and lower background to approach the zero background condition.

\begin{acknowledgments}
The authors acknowledge the help and support from UCIL, Jaduguda Mine for the execution of the experiment at JUSL. The authors are grateful to Professor Satyaki Bhattacharya, SINP for the helpful discussion in the statistical analysis process, Professor Maitreyee Nandy, SINP for the simulation with FLUKA and Professor Anirban Das, SINP for the useful discussion on theoretical models. The authors are thankful to the mechanical workshop, SINP for the experimental setup. The authors are also thankful to HPU, VECC for providing the \(^{241}\)AmBe source. The work is supported by the DAE, GoI, SINP/4001 project fund. V. K. is thankful to CSIR for the support of his fellowship.
\end{acknowledgments}

\section*{DATA AVAILABILITY}
The data that support the findings of this article are not publicly available. The data are available from the authors upon reasonable request.


\bibliography{apssamp}  

\begin{thebibliography}{41}%
\makeatletter
\providecommand \@ifxundefined [1]{%
 \@ifx{#1\undefined}
}%
\providecommand \@ifnum [1]{%
 \ifnum #1\expandafter \@firstoftwo
 \else \expandafter \@secondoftwo
 \fi
}%
\providecommand \@ifx [1]{%
 \ifx #1\expandafter \@firstoftwo
 \else \expandafter \@secondoftwo
 \fi
}%
\providecommand \natexlab [1]{#1}%
\providecommand \enquote  [1]{``#1''}%
\providecommand \bibnamefont  [1]{#1}%
\providecommand \bibfnamefont [1]{#1}%
\providecommand \citenamefont [1]{#1}%
\providecommand \href@noop [0]{\@secondoftwo}%
\providecommand \href [0]{\begingroup \@sanitize@url \@href}%
\providecommand \@href[1]{\@@startlink{#1}\@@href}%
\providecommand \@@href[1]{\endgroup#1\@@endlink}%
\providecommand \@sanitize@url [0]{\catcode `\\12\catcode `\$12\catcode `\&12\catcode `\#12\catcode `\^12\catcode `\_12\catcode `\%12\relax}%
\providecommand \@@startlink[1]{}%
\providecommand \@@endlink[0]{}%
\providecommand \url  [0]{\begingroup\@sanitize@url \@url }%
\providecommand \@url [1]{\endgroup\@href {#1}{\urlprefix }}%
\providecommand \urlprefix  [0]{URL }%
\providecommand \Eprint [0]{\href }%
\providecommand \doibase [0]{https://doi.org/}%
\providecommand \selectlanguage [0]{\@gobble}%
\providecommand \bibinfo  [0]{\@secondoftwo}%
\providecommand \bibfield  [0]{\@secondoftwo}%
\providecommand \translation [1]{[#1]}%
\providecommand \BibitemOpen [0]{}%
\providecommand \bibitemStop [0]{}%
\providecommand \bibitemNoStop [0]{.\EOS\space}%
\providecommand \EOS [0]{\spacefactor3000\relax}%
\providecommand \BibitemShut  [1]{\csname bibitem#1\endcsname}%
\let\auto@bib@innerbib\@empty
\bibitem [{\citenamefont {D'Amico}\ \emph {et~al.}()\citenamefont {D'Amico}, \citenamefont {Kamionkowski},\ and\ \citenamefont {Sigurdson}}]{damico2009darkmatterastrophysics}%
  \BibitemOpen
  \bibfield  {author} {\bibinfo {author} {\bibfnamefont {G.}~\bibnamefont {D'Amico}}, \bibinfo {author} {\bibfnamefont {M.}~\bibnamefont {Kamionkowski}},\ and\ \bibinfo {author} {\bibfnamefont {K.}~\bibnamefont {Sigurdson}},\ }\href {https://arxiv.org/abs/0907.1912} {}\Eprint {https://arxiv.org/abs/0907.1912} {arXiv:0907.1912} \BibitemShut {NoStop}%
\bibitem [{\citenamefont {Roszkowski}\ \emph {et~al.}(2018)\citenamefont {Roszkowski}, \citenamefont {Sessolo},\ and\ \citenamefont {Trojanowski}}]{Roszkowski_2018}%
  \BibitemOpen
  \bibfield  {author} {\bibinfo {author} {\bibfnamefont {L.}~\bibnamefont {Roszkowski}}, \bibinfo {author} {\bibfnamefont {E.~M.}\ \bibnamefont {Sessolo}},\ and\ \bibinfo {author} {\bibfnamefont {S.}~\bibnamefont {Trojanowski}},\ }\href {https://doi.org/10.1088/1361-6633/aab913} {\bibfield  {journal} {\bibinfo  {journal} {Rep. Prog. Phys.}\ }\textbf {\bibinfo {volume} {81}},\ \bibinfo {pages} {066201} (\bibinfo {year} {2018})}\BibitemShut {NoStop}%
\bibitem [{\citenamefont {Zhang}\ \emph {et~al.}(2022)\citenamefont {Zhang} \emph {et~al.}}]{PhysRevLett.129.221301}%
  \BibitemOpen
  \bibfield  {author} {\bibinfo {author} {\bibfnamefont {Z.~Y.}\ \bibnamefont {Zhang}} \emph {et~al.} (\bibinfo {collaboration} {CDEX Collaboration}),\ }\href {https://doi.org/10.1103/PhysRevLett.129.221301} {\bibfield  {journal} {\bibinfo  {journal} {Phys. Rev. Lett.}\ }\textbf {\bibinfo {volume} {129}},\ \bibinfo {pages} {221301} (\bibinfo {year} {2022})}\BibitemShut {NoStop}%
\bibitem [{\citenamefont {Aprile}\ \emph {et~al.}(2023)\citenamefont {Aprile} \emph {et~al.}}]{PhysRevLett.131.041003}%
  \BibitemOpen
  \bibfield  {author} {\bibinfo {author} {\bibfnamefont {E.}~\bibnamefont {Aprile}} \emph {et~al.} (\bibinfo {collaboration} {XENON Collaboration}),\ }\href {https://doi.org/10.1103/PhysRevLett.131.041003} {\bibfield  {journal} {\bibinfo  {journal} {Phys. Rev. Lett.}\ }\textbf {\bibinfo {volume} {131}},\ \bibinfo {pages} {041003} (\bibinfo {year} {2023})}\BibitemShut {NoStop}%
\bibitem [{\citenamefont {Aalbers}\ \emph {et~al.}(2023)\citenamefont {Aalbers} \emph {et~al.}}]{PhysRevLett.131.041002}%
  \BibitemOpen
  \bibfield  {author} {\bibinfo {author} {\bibfnamefont {J.}~\bibnamefont {Aalbers}} \emph {et~al.} (\bibinfo {collaboration} {LUX-ZEPLIN Collaboration}),\ }\href {https://doi.org/10.1103/PhysRevLett.131.041002} {\bibfield  {journal} {\bibinfo  {journal} {Phys. Rev. Lett.}\ }\textbf {\bibinfo {volume} {131}},\ \bibinfo {pages} {041002} (\bibinfo {year} {2023})}\BibitemShut {NoStop}%
\bibitem [{\citenamefont {Li}\ \emph {et~al.}(2023)\citenamefont {Li} \emph {et~al.}}]{PhysRevLett.130.261001}%
  \BibitemOpen
  \bibfield  {author} {\bibinfo {author} {\bibfnamefont {S.}~\bibnamefont {Li}} \emph {et~al.} (\bibinfo {collaboration} {PandaX Collaboration}),\ }\href {https://doi.org/10.1103/PhysRevLett.130.261001} {\bibfield  {journal} {\bibinfo  {journal} {Phys. Rev. Lett.}\ }\textbf {\bibinfo {volume} {130}},\ \bibinfo {pages} {261001} (\bibinfo {year} {2023})}\BibitemShut {NoStop}%
\bibitem [{\citenamefont {Arnaud}\ \emph {et~al.}(2018)\citenamefont {Arnaud} \emph {et~al.}}]{ARNAUD201854}%
  \BibitemOpen
  \bibfield  {author} {\bibinfo {author} {\bibfnamefont {Q.}~\bibnamefont {Arnaud}} \emph {et~al.},\ }\href {https://doi.org/https://doi.org/10.1016/j.astropartphys.2017.10.009} {\bibfield  {journal} {\bibinfo  {journal} {Astropart. Phys.}\ }\textbf {\bibinfo {volume} {97}},\ \bibinfo {pages} {54} (\bibinfo {year} {2018})}\BibitemShut {NoStop}%
\bibitem [{\citenamefont {Agnes}\ \emph {et~al.}(2015)\citenamefont {Agnes} \emph {et~al.}}]{AGNES2015456}%
  \BibitemOpen
  \bibfield  {author} {\bibinfo {author} {\bibfnamefont {P.}~\bibnamefont {Agnes}} \emph {et~al.},\ }\href {https://doi.org/https://doi.org/10.1016/j.physletb.2015.03.012} {\bibfield  {journal} {\bibinfo  {journal} {Phys. Lett. B}\ }\textbf {\bibinfo {volume} {743}},\ \bibinfo {pages} {456} (\bibinfo {year} {2015})}\BibitemShut {NoStop}%
\bibitem [{\citenamefont {Behnke}\ \emph {et~al.}(2017)\citenamefont {Behnke} \emph {et~al.}}]{BEHNKE201785}%
  \BibitemOpen
  \bibfield  {author} {\bibinfo {author} {\bibfnamefont {E.}~\bibnamefont {Behnke}} \emph {et~al.},\ }\href {https://doi.org/https://doi.org/10.1016/j.astropartphys.2017.02.005} {\bibfield  {journal} {\bibinfo  {journal} {Astropart. Phys.}\ }\textbf {\bibinfo {volume} {90}},\ \bibinfo {pages} {85} (\bibinfo {year} {2017})}\BibitemShut {NoStop}%
\bibitem [{\citenamefont {Amole}\ \emph {et~al.}(2019)\citenamefont {Amole} \emph {et~al.}}]{PhysRevD.100.022001}%
  \BibitemOpen
  \bibfield  {author} {\bibinfo {author} {\bibfnamefont {C.}~\bibnamefont {Amole}} \emph {et~al.} (\bibinfo {collaboration} {PICO Collaboration}),\ }\href {https://doi.org/10.1103/PhysRevD.100.022001} {\bibfield  {journal} {\bibinfo  {journal} {Phys. Rev. D}\ }\textbf {\bibinfo {volume} {100}},\ \bibinfo {pages} {022001} (\bibinfo {year} {2019})}\BibitemShut {NoStop}%
\bibitem [{\citenamefont {Felizardo}\ \emph {et~al.}(2012)\citenamefont {Felizardo} \emph {et~al.}}]{PhysRevLett.108.201302}%
  \BibitemOpen
  \bibfield  {author} {\bibinfo {author} {\bibfnamefont {M.}~\bibnamefont {Felizardo}} \emph {et~al.} (\bibinfo {collaboration} {The SIMPLE Collaboration}),\ }\href {https://doi.org/10.1103/PhysRevLett.108.201302} {\bibfield  {journal} {\bibinfo  {journal} {Phys. Rev. Lett.}\ }\textbf {\bibinfo {volume} {108}},\ \bibinfo {pages} {201302} (\bibinfo {year} {2012})}\BibitemShut {NoStop}%
\bibitem [{\citenamefont {Antonicci}\ \emph {et~al.}(2017)\citenamefont {Antonicci} \emph {et~al.}}]{Antonicci2017}%
  \BibitemOpen
  \bibfield  {author} {\bibinfo {author} {\bibfnamefont {A.}~\bibnamefont {Antonicci}} \emph {et~al.} (\bibinfo {collaboration} {The MOSCAB Collaboration}),\ }\href {https://doi.org/10.1140/epjc/s10052-017-5313-8} {\bibfield  {journal} {\bibinfo  {journal} {Eur. Phys. J. C}\ }\textbf {\bibinfo {volume} {77}},\ \bibinfo {pages} {752} (\bibinfo {year} {2017})}\BibitemShut {NoStop}%
\bibitem [{\citenamefont {Fustin}()}]{fustin2024darkmatterlimitscoupp}%
  \BibitemOpen
  \bibfield  {author} {\bibinfo {author} {\bibfnamefont {D.}~\bibnamefont {Fustin}},\ }\href {https://arxiv.org/abs/2401.07384} {}\Eprint {https://arxiv.org/abs/2401.07384} {arXiv:2401.07384} \BibitemShut {NoStop}%
\bibitem [{\citenamefont {Angloher}\ \emph {et~al.}(2023)\citenamefont {Angloher} \emph {et~al.}}]{PhysRevD.107.122003}%
  \BibitemOpen
  \bibfield  {author} {\bibinfo {author} {\bibfnamefont {G.}~\bibnamefont {Angloher}} \emph {et~al.} (\bibinfo {collaboration} {CRESST Collaboration}),\ }\href {https://doi.org/10.1103/PhysRevD.107.122003} {\bibfield  {journal} {\bibinfo  {journal} {Phys. Rev. D}\ }\textbf {\bibinfo {volume} {107}},\ \bibinfo {pages} {122003} (\bibinfo {year} {2023})}\BibitemShut {NoStop}%
\bibitem [{\citenamefont {Agnese}\ \emph {et~al.}(2019)\citenamefont {Agnese} \emph {et~al.}}]{PhysRevD.99.062001}%
  \BibitemOpen
  \bibfield  {author} {\bibinfo {author} {\bibfnamefont {R.}~\bibnamefont {Agnese}} \emph {et~al.} (\bibinfo {collaboration} {SuperCDMS Collaboration}),\ }\href {https://doi.org/10.1103/PhysRevD.99.062001} {\bibfield  {journal} {\bibinfo  {journal} {Phys. Rev. D}\ }\textbf {\bibinfo {volume} {99}},\ \bibinfo {pages} {062001} (\bibinfo {year} {2019})}\BibitemShut {NoStop}%
\bibitem [{\citenamefont {Jiang}\ \emph {et~al.}(2018)\citenamefont {Jiang} \emph {et~al.}}]{PhysRevLett.120.241301}%
  \BibitemOpen
  \bibfield  {author} {\bibinfo {author} {\bibfnamefont {H.}~\bibnamefont {Jiang}} \emph {et~al.} (\bibinfo {collaboration} {CDEX Collaboration}),\ }\href {https://doi.org/10.1103/PhysRevLett.120.241301} {\bibfield  {journal} {\bibinfo  {journal} {Phys. Rev. Lett.}\ }\textbf {\bibinfo {volume} {120}},\ \bibinfo {pages} {241301} (\bibinfo {year} {2018})}\BibitemShut {NoStop}%
\bibitem [{\citenamefont {Liu}\ \emph {et~al.}(2019)\citenamefont {Liu} \emph {et~al.}}]{PhysRevLett.123.161301}%
  \BibitemOpen
  \bibfield  {author} {\bibinfo {author} {\bibfnamefont {Z.~Z.}\ \bibnamefont {Liu}} \emph {et~al.} (\bibinfo {collaboration} {CDEX Collaboration}),\ }\href {https://doi.org/10.1103/PhysRevLett.123.161301} {\bibfield  {journal} {\bibinfo  {journal} {Phys. Rev. Lett.}\ }\textbf {\bibinfo {volume} {123}},\ \bibinfo {pages} {161301} (\bibinfo {year} {2019})}\BibitemShut {NoStop}%
\bibitem [{\citenamefont {Akerib}\ \emph {et~al.}(2019)\citenamefont {Akerib} \emph {et~al.}}]{PhysRevLett.122.131301}%
  \BibitemOpen
  \bibfield  {author} {\bibinfo {author} {\bibfnamefont {D.~S.}\ \bibnamefont {Akerib}} \emph {et~al.} (\bibinfo {collaboration} {LUX Collaboration}),\ }\href {https://doi.org/10.1103/PhysRevLett.122.131301} {\bibfield  {journal} {\bibinfo  {journal} {Phys. Rev. Lett.}\ }\textbf {\bibinfo {volume} {122}},\ \bibinfo {pages} {131301} (\bibinfo {year} {2019})}\BibitemShut {NoStop}%
\bibitem [{\citenamefont {Angloher}\ \emph {et~al.}(2017)\citenamefont {Angloher} \emph {et~al.}}]{Angloher2017}%
  \BibitemOpen
  \bibfield  {author} {\bibinfo {author} {\bibfnamefont {G.}~\bibnamefont {Angloher}} \emph {et~al.},\ }\href {https://doi.org/10.1140/epjc/s10052-017-5223-9} {\bibfield  {journal} {\bibinfo  {journal} {Eur. Phys. J. C}\ }\textbf {\bibinfo {volume} {77}},\ \bibinfo {pages} {637} (\bibinfo {year} {2017})}\BibitemShut {NoStop}%
\bibitem [{\citenamefont {Adari}\ \emph {et~al.}(2025)\citenamefont {Adari} \emph {et~al.}}]{PhysRevLett.134.011804}%
  \BibitemOpen
  \bibfield  {author} {\bibinfo {author} {\bibfnamefont {P.}~\bibnamefont {Adari}} \emph {et~al.} (\bibinfo {collaboration} {SENSEI Collaboration}),\ }\href {https://doi.org/10.1103/PhysRevLett.134.011804} {\bibfield  {journal} {\bibinfo  {journal} {Phys. Rev. Lett.}\ }\textbf {\bibinfo {volume} {134}},\ \bibinfo {pages} {011804} (\bibinfo {year} {2025})}\BibitemShut {NoStop}%
\bibitem [{\citenamefont {Aguilar-Arevalo}\ \emph {et~al.}(2019)\citenamefont {Aguilar-Arevalo} \emph {et~al.}}]{PhysRevLett.123.181802}%
  \BibitemOpen
  \bibfield  {author} {\bibinfo {author} {\bibfnamefont {A.}~\bibnamefont {Aguilar-Arevalo}} \emph {et~al.} (\bibinfo {collaboration} {DAMIC Collaboration}),\ }\href {https://doi.org/10.1103/PhysRevLett.123.181802} {\bibfield  {journal} {\bibinfo  {journal} {Phys. Rev. Lett.}\ }\textbf {\bibinfo {volume} {123}},\ \bibinfo {pages} {181802} (\bibinfo {year} {2019})}\BibitemShut {NoStop}%
\bibitem [{\citenamefont {Arnaud}\ \emph {et~al.}(2020)\citenamefont {Arnaud} \emph {et~al.}}]{PhysRevLett.125.141301}%
  \BibitemOpen
  \bibfield  {author} {\bibinfo {author} {\bibfnamefont {Q.}~\bibnamefont {Arnaud}} \emph {et~al.} (\bibinfo {collaboration} {EDELWEISS Collaboration}),\ }\href {https://doi.org/10.1103/PhysRevLett.125.141301} {\bibfield  {journal} {\bibinfo  {journal} {Phys. Rev. Lett.}\ }\textbf {\bibinfo {volume} {125}},\ \bibinfo {pages} {141301} (\bibinfo {year} {2020})}\BibitemShut {NoStop}%
\bibitem [{\citenamefont {Amaral}\ \emph {et~al.}(2020)\citenamefont {Amaral} \emph {et~al.}}]{PhysRevD.102.091101}%
  \BibitemOpen
  \bibfield  {author} {\bibinfo {author} {\bibfnamefont {D.~W.}\ \bibnamefont {Amaral}} \emph {et~al.},\ }\href {https://doi.org/10.1103/PhysRevD.102.091101} {\bibfield  {journal} {\bibinfo  {journal} {Phys. Rev. D}\ }\textbf {\bibinfo {volume} {102}},\ \bibinfo {pages} {091101} (\bibinfo {year} {2020})}\BibitemShut {NoStop}%
\bibitem [{\citenamefont {Essig}\ \emph {et~al.}(2017)\citenamefont {Essig}, \citenamefont {Volansky},\ and\ \citenamefont {Yu}}]{PhysRevD.96.043017}%
  \BibitemOpen
  \bibfield  {author} {\bibinfo {author} {\bibfnamefont {R.}~\bibnamefont {Essig}}, \bibinfo {author} {\bibfnamefont {T.}~\bibnamefont {Volansky}},\ and\ \bibinfo {author} {\bibfnamefont {T.-T.}\ \bibnamefont {Yu}},\ }\href {https://doi.org/10.1103/PhysRevD.96.043017} {\bibfield  {journal} {\bibinfo  {journal} {Phys. Rev. D}\ }\textbf {\bibinfo {volume} {96}},\ \bibinfo {pages} {043017} (\bibinfo {year} {2017})}\BibitemShut {NoStop}%
\bibitem [{\citenamefont {Aprile}\ \emph {et~al.}(2019)\citenamefont {Aprile} \emph {et~al.}}]{PhysRevLett.123.251801}%
  \BibitemOpen
  \bibfield  {author} {\bibinfo {author} {\bibfnamefont {E.}~\bibnamefont {Aprile}} \emph {et~al.} (\bibinfo {collaboration} {XENON Collaboration}),\ }\href {https://doi.org/10.1103/PhysRevLett.123.251801} {\bibfield  {journal} {\bibinfo  {journal} {Phys. Rev. Lett.}\ }\textbf {\bibinfo {volume} {123}},\ \bibinfo {pages} {251801} (\bibinfo {year} {2019})}\BibitemShut {NoStop}%
\bibitem [{\citenamefont {Cheng}\ \emph {et~al.}(2021)\citenamefont {Cheng} \emph {et~al.}}]{PhysRevLett.126.211803}%
  \BibitemOpen
  \bibfield  {author} {\bibinfo {author} {\bibfnamefont {C.}~\bibnamefont {Cheng}} \emph {et~al.} (\bibinfo {collaboration} {PandaX-II Collaboration}),\ }\href {https://doi.org/10.1103/PhysRevLett.126.211803} {\bibfield  {journal} {\bibinfo  {journal} {Phys. Rev. Lett.}\ }\textbf {\bibinfo {volume} {126}},\ \bibinfo {pages} {211803} (\bibinfo {year} {2021})}\BibitemShut {NoStop}%
\bibitem [{\citenamefont {Agnes}\ \emph {et~al.}(2018)\citenamefont {Agnes} \emph {et~al.}}]{PhysRevLett.121.111303}%
  \BibitemOpen
  \bibfield  {author} {\bibinfo {author} {\bibfnamefont {P.}~\bibnamefont {Agnes}} \emph {et~al.} (\bibinfo {collaboration} {The DarkSide Collaboration}),\ }\href {https://doi.org/10.1103/PhysRevLett.121.111303} {\bibfield  {journal} {\bibinfo  {journal} {Phys. Rev. Lett.}\ }\textbf {\bibinfo {volume} {121}},\ \bibinfo {pages} {111303} (\bibinfo {year} {2018})}\BibitemShut {NoStop}%
\bibitem [{\citenamefont {Seth}\ \emph {et~al.}(2020)\citenamefont {Seth}, \citenamefont {Sahoo}, \citenamefont {Bhattacharjee},\ and\ \citenamefont {Das}}]{PhysRevD.101.103005}%
  \BibitemOpen
  \bibfield  {author} {\bibinfo {author} {\bibfnamefont {S.}~\bibnamefont {Seth}}, \bibinfo {author} {\bibfnamefont {S.}~\bibnamefont {Sahoo}}, \bibinfo {author} {\bibfnamefont {P.}~\bibnamefont {Bhattacharjee}},\ and\ \bibinfo {author} {\bibfnamefont {M.}~\bibnamefont {Das}},\ }\href {https://doi.org/10.1103/PhysRevD.101.103005} {\bibfield  {journal} {\bibinfo  {journal} {Phys. Rev. D}\ }\textbf {\bibinfo {volume} {101}},\ \bibinfo {pages} {103005} (\bibinfo {year} {2020})}\BibitemShut {NoStop}%
\bibitem [{\citenamefont {Ghosh}\ \emph {et~al.}(2022)\citenamefont {Ghosh}, \citenamefont {Dutta}, \citenamefont {Mondal},\ and\ \citenamefont {Saha}}]{GHOSH2022102700}%
  \BibitemOpen
  \bibfield  {author} {\bibinfo {author} {\bibfnamefont {S.}~\bibnamefont {Ghosh}}, \bibinfo {author} {\bibfnamefont {S.}~\bibnamefont {Dutta}}, \bibinfo {author} {\bibfnamefont {N.~K.}\ \bibnamefont {Mondal}},\ and\ \bibinfo {author} {\bibfnamefont {S.}~\bibnamefont {Saha}},\ }\href {https://doi.org/https://doi.org/10.1016/j.astropartphys.2022.102700} {\bibfield  {journal} {\bibinfo  {journal} {Astropart. Phys.}\ }\textbf {\bibinfo {volume} {139}},\ \bibinfo {pages} {102700} (\bibinfo {year} {2022})}\BibitemShut {NoStop}%
\bibitem [{\citenamefont {Kumar}\ \emph {et~al.}()\citenamefont {Kumar}, \citenamefont {Ali}, \citenamefont {Das}, \citenamefont {Biswas}, \citenamefont {Das}, \citenamefont {Sahoo}, \citenamefont {Chaddha}, \citenamefont {Basu},\ and\ \citenamefont {Jha}}]{kumar2025resulttetrafluoroethanec2h2f4superheated}%
  \BibitemOpen
  \bibfield  {author} {\bibinfo {author} {\bibfnamefont {V.}~\bibnamefont {Kumar}}, \bibinfo {author} {\bibfnamefont {S.}~\bibnamefont {Ali}}, \bibinfo {author} {\bibfnamefont {M.}~\bibnamefont {Das}}, \bibinfo {author} {\bibfnamefont {N.}~\bibnamefont {Biswas}}, \bibinfo {author} {\bibfnamefont {S.}~\bibnamefont {Das}}, \bibinfo {author} {\bibfnamefont {S.}~\bibnamefont {Sahoo}}, \bibinfo {author} {\bibfnamefont {N.}~\bibnamefont {Chaddha}}, \bibinfo {author} {\bibfnamefont {J.}~\bibnamefont {Basu}},\ and\ \bibinfo {author} {\bibfnamefont {V.~N.}\ \bibnamefont {Jha}},\ }\href {https://arxiv.org/abs/2501.02833} {}\Eprint {https://arxiv.org/abs/2501.02833} {arXiv:2501.02833} \BibitemShut {NoStop}%
\bibitem [{\citenamefont {Sahoo}\ \emph {et~al.}(2021{\natexlab{a}})\citenamefont {Sahoo} \emph {et~al.}}]{SAHOO2021165457}%
  \BibitemOpen
  \bibfield  {author} {\bibinfo {author} {\bibfnamefont {S.}~\bibnamefont {Sahoo}} \emph {et~al.},\ }\href {https://doi.org/https://doi.org/10.1016/j.nima.2021.165457} {\bibfield  {journal} {\bibinfo  {journal} {Nucl. Instrum. Methods Phys. Res. Sect. A}\ }\textbf {\bibinfo {volume} {1009}},\ \bibinfo {pages} {165457} (\bibinfo {year} {2021}{\natexlab{a}})}\BibitemShut {NoStop}%
\bibitem [{\citenamefont {Sahoo}\ \emph {et~al.}(2021{\natexlab{b}})\citenamefont {Sahoo} \emph {et~al.}}]{SAHOO2021165450}%
  \BibitemOpen
  \bibfield  {author} {\bibinfo {author} {\bibfnamefont {S.}~\bibnamefont {Sahoo}} \emph {et~al.},\ }\href {https://doi.org/https://doi.org/10.1016/j.nima.2021.165450} {\bibfield  {journal} {\bibinfo  {journal} {Nucl. Instrum. Methods Phys. Res. Sect. A}\ }\textbf {\bibinfo {volume} {1008}},\ \bibinfo {pages} {165450} (\bibinfo {year} {2021}{\natexlab{b}})}\BibitemShut {NoStop}%
\bibitem [{\citenamefont {Sahoo}\ \emph {et~al.}(2019)\citenamefont {Sahoo}, \citenamefont {Seth},\ and\ \citenamefont {Das}}]{SAHOO201944}%
  \BibitemOpen
  \bibfield  {author} {\bibinfo {author} {\bibfnamefont {S.}~\bibnamefont {Sahoo}}, \bibinfo {author} {\bibfnamefont {S.}~\bibnamefont {Seth}},\ and\ \bibinfo {author} {\bibfnamefont {M.}~\bibnamefont {Das}},\ }\href {https://doi.org/https://doi.org/10.1016/j.nima.2019.04.010} {\bibfield  {journal} {\bibinfo  {journal} {Nucl. Instrum. Methods Phys. Res. Sect. A}\ }\textbf {\bibinfo {volume} {931}},\ \bibinfo {pages} {44} (\bibinfo {year} {2019})}\BibitemShut {NoStop}%
\bibitem [{\citenamefont {Zeigler}\ \emph {et~al.}()\citenamefont {Zeigler} \emph {et~al.}}]{Zeigler}%
  \BibitemOpen
  \bibfield  {author} {\bibinfo {author} {\bibfnamefont {J.~F.}\ \bibnamefont {Zeigler}} \emph {et~al.},\ }\href@noop {} {}\bibinfo {howpublished} {http://www.srim.org.}\BibitemShut {Stop}%
\bibitem [{flu()}]{fluka}%
  \BibitemOpen
  \href@noop {} {}\bibinfo {howpublished} {https://fluka.cern}\BibitemShut {NoStop}%
\bibitem [{fla()}]{flair}%
  \BibitemOpen
  \href@noop {} {}\bibinfo {howpublished} {http://cern.ch/flair}\BibitemShut {NoStop}%
\bibitem [{NND()}]{NNDC}%
  \BibitemOpen
  \href@noop {} {}\bibinfo {howpublished} {https://www.nndc.bnl.gov}\BibitemShut {NoStop}%
\bibitem [{\citenamefont {Cowan}\ \emph {et~al.}(2013)\citenamefont {Cowan}, \citenamefont {Cranmer}, \citenamefont {Gross},\ and\ \citenamefont {Vitells}}]{article}%
  \BibitemOpen
  \bibfield  {author} {\bibinfo {author} {\bibfnamefont {G.}~\bibnamefont {Cowan}}, \bibinfo {author} {\bibfnamefont {K.}~\bibnamefont {Cranmer}}, \bibinfo {author} {\bibfnamefont {E.}~\bibnamefont {Gross}},\ and\ \bibinfo {author} {\bibfnamefont {O.}~\bibnamefont {Vitells}},\ }\href {https://doi.org/10.1140/epjc/s10052-013-2501-z} {\bibfield  {journal} {\bibinfo  {journal} {Eur. Phys. J. C}\ }\textbf {\bibinfo {volume} {73}},\ \bibinfo {pages} {2501} (\bibinfo {year} {2013})}\BibitemShut {NoStop}%
\bibitem [{\citenamefont {Baxter}\ \emph {et~al.}(2021)\citenamefont {Baxter} \emph {et~al.}}]{Baxter2021}%
  \BibitemOpen
  \bibfield  {author} {\bibinfo {author} {\bibfnamefont {D.}~\bibnamefont {Baxter}} \emph {et~al.},\ }\href {https://doi.org/10.1140/epjc/s10052-021-09655-y} {\bibfield  {journal} {\bibinfo  {journal} {Eur. Phys. J. C}\ }\textbf {\bibinfo {volume} {81}},\ \bibinfo {pages} {907} (\bibinfo {year} {2021})}\BibitemShut {NoStop}%
\bibitem [{\citenamefont {Essig}\ \emph {et~al.}()\citenamefont {Essig} \emph {et~al.}}]{essig2023snowmass2021cosmicfrontierlandscape}%
  \BibitemOpen
  \bibfield  {author} {\bibinfo {author} {\bibfnamefont {R.}~\bibnamefont {Essig}} \emph {et~al.},\ }\href {https://arxiv.org/abs/2203.08297} {}\Eprint {https://arxiv.org/abs/2203.08297} {arXiv:2203.08297} \BibitemShut {NoStop}%
\bibitem [{\citenamefont {Battaglieri}\ \emph {et~al.}()\citenamefont {Battaglieri} \emph {et~al.}}]{battaglieri2017cosmicvisionsnewideas}%
  \BibitemOpen
  \bibfield  {author} {\bibinfo {author} {\bibfnamefont {M.}~\bibnamefont {Battaglieri}} \emph {et~al.},\ }\href {https://arxiv.org/abs/1707.04591} {}\Eprint {https://arxiv.org/abs/1707.04591} {arXiv:1707.04591} \BibitemShut {NoStop}%
\end{thebibliography}%
\end{document}